\documentclass[twocolumn,showpacs,preprintnumbers,showkeys,superscriptaddress]{revtex4}

\usepackage{bbm}
\usepackage{amssymb}
\usepackage{tipa}
\usepackage{mathrsfs}
\usepackage{amssymb}
\usepackage{graphicx}
\usepackage{dcolumn}
\usepackage{bm}
\usepackage{graphicx}

\newfont{\largemi}{cmmi10}
\newfont{\smallmi}{cmmi6}

\draft

\def\eqref#1{Eq.~(\ref{eq:#1})}

\begin{document}

\title{Spherical to deformed shape transition in  the nucleon-pair shell model}

\author{Y. Lei}

\affiliation{INPAC, Department of Physics and Shanghai Key
  Lab for Particle Physics and Cosmology, Shanghai Jiao Tong University,
 Shanghai, 200240, China}
\affiliation{Bartol Research Institute and Department of Physics
and Astronomy, University of Delaware, Newark, Delaware 19716, USA}

\author{S. Pittel}
\affiliation{Bartol Research Institute and Department of Physics and Astronomy,
University of Delaware, Newark, Delaware 19716, USA}

\author{G. J. Fu}
\affiliation{INPAC, Department of Physics and Shanghai Key
 Lab for Particle Physics and Cosmology, Shanghai Jiao Tong University,
 Shanghai, 200240, China}
\affiliation{Bartol Research Institute and Department of Physics
and Astronomy, University of Delaware, Newark, Delaware 19716, USA}

\author{Y. M. Zhao }
\affiliation{INPAC, Department of Physics and Shanghai Key
  Lab for Particle Physics and Cosmology, Shanghai Jiao Tong University,
  Shanghai, 200240, China}  \affiliation{Center of Theoretical
Nuclear Physics, National Laboratory of Heavy Ion Accelerator, Lanzhou 730000, China}

\date{\today}

\begin{abstract}
A study of the shape transition from spherical to axially deformed
nuclei in the even Ce isotopes using the nucleon-pair approximation
of the shell model  is reported.  As long as the structure of the
dominant collective pairs is determined using a microscopic
framework appropriate to deformed nuclei, the model is able to
produce a shape transition. However, the resulting transition is too
rapid, with nuclei that should be transitional being fairly well
deformed, perhaps reflecting the need to maintain several pairs with
each angular momentum.
\end{abstract}
\maketitle

\section{Introduction}  \label{int}

One of the most fascinating aspects of nuclear structure is the
existence of competing modes of collective motion, with phase
transitions linking them. This has been discussed extensively within
the context of the interacting boson model (IBM) \cite{IBM}, with
the vertices of the Casten triangle representing extreme modes of
collective motion in  the ground state \cite{phase}.

There has been much effort to develop theories that are able to describe the
competing modes of nuclear collectivity and the associated transitions from one
to another. This is a major theme of the ongoing effort to develop a unified
density functional theory of atomic nuclei. It has also been a major theme of
the phenomenological IBM, which has recently seen a significant microscopic
advance through the development of a mapping procedure that translates the
physics of density functionals to the IBM parameters in the various collective
regions \cite{Otsuka}.

A model that is closely linked in spirit to the IBM is the
nucleon-pair approximation (NPA)  of the shell model (often called
the nucleon-pair shell model in the literature) \cite{Chen,NPSM}.
Like the IBM, the NPA  is based on a truncation in terms of pair
degrees of freedom. Unlike the IBM, however, the microscopic
structure of the pairs is preserved throughout.

Recently, it has been shown that the NPA truncated to the dominant
$S$ and $D$ pair degrees of freedom is able to qualitatively
reproduce all of the phase transitions of the IBM, including the
transition from vibrational U(5) behavior to axially symmetric SU(3)
rotational behavior \cite{Draayer}. These studies were of a model
nature, however, not making contact with real nuclear systems.

In this work, we return to the issue of whether the NPA is able to describe the
shape transition from vibrational to rotational behavior in real nuclei. We
carry out the analysis for the even-$A$ Ce isotopes from $^{142}$Ce through
$^{148}$Ce, where such a shape transition has been observed experimentally. We
find that when we use the traditional approach for generating the dominant
collective pairs of the model, we cannot reproduce the shape transition. We
then suggest the use of an alternative approach to the structure of these pairs
for rotational nuclei that incorporates more of the physics that builds their
correlations. With this new approach to the structure of the collective pairs
in the rotational region, we are able to achieve the observed shape transition.

The structure of this article is as follows. In Sec. \ref{npa}, we briefly
review the basic ingredients of the NPA, including a
discussion of the traditional approach for defining the structure of the
collective pairs to be used when applying the method. In Sec. \ref{hfb},
we describe the alternative formalism for building the collective pairs
appropriate to deformed nuclei. In Sec. \ref{ham}, we discuss other input to
our calculations for the even-mass Ce isotopes, for which the results are
presented in Sec. \ref{res}. Finally, in Sec. \ref{sum} we summarize the key
conclusions of the work and outline some issues for future consideration.

\section{The nucleon-pair approximation of the shell model}\label{npa}

The nucleon-pair approximation of the shell model (NPA) is a truncation
strategy for the nuclear shell model based on the dominance of a few
selected coherent pair configurations. When viewed as a microscopic
realization of the IBM, the selected nucleon pairs are
the lowest spin-zero and spin-two pairs for neutrons and
protons, called $S$ and $D$ pairs, respectively. The NPA does not map these collective pairs onto bosons,
however, working in terms of the collective pair configurations
directly. Furthermore, the NPA does not restrict the collective
pairs that are included to the lowest spin-zero and spin-two pair configurations for neutrons
and protons only, permitting in principle other pair configurations to be included.

The NPA traces back to the generalized seniority scheme
\cite{GeneralizedSeniority} and the broken pair approximation (BPA)
\cite{BPA}. These methods could not be used as a practical and
general truncation strategy for the shell model, however,  until the
development by Chen \cite{Chen} in 1997 of a set of recursive
formulae that enabled the calculation of the matrix elements of {\it
realistic} nuclear hamiltonian in a basis built up in terms of these
collective pairs. We italicize the term {\it realistic} because of
the computational limitation to date of the methodology to
hamiltonians involving multipole pairing interactions in the
like-nucleon channels and separable multipole-multipole interactions
in the neutron-proton channel. With such assumed semi-realistic
hamiltonians and codes based on the Chen's recursion formulae, it is
now possible to treat systems involving several pairs of each type
of particle and several collective pair degrees of freedom.

There are several key steps in the NPA strategy. One involves a
choice of the effective shell-model hamiltonian, along the lines
just discussed. The second involves an assumption of the pair degrees
of freedom to be included, as dictated by the physics of the problem
under discussion and computational limitations. Another key
ingredient is the assumed structure of the collective pairs that are
retained. The standard approach has been to obtain the lowest
$S$ pair from a number-conserving BCS treatment, or
equivalently a variational description in terms of states with
generalized seniority zero, for neutrons and protons, respectively.
The prescription typically used for the lowest excited pairs is a
variational treatment of these pairs in terms of generalized
seniority-two states, using the $S$ pairs already determined. An
alternative approach sometimes used for the structure of the $D$
pair is to generate it from the commutator of the quadrupole
operator with the $S$ pair. The traditional prescription is
reasonable for systems that are meaningfully described by the
generalized seniority or broken pair scheme, namely for either
semi-magic or near semi-magic nuclei. Once there are a large enough
number of both types of particles, however, the expectation is that
generalized seniority ceases to provide a useful description of the
states of the system. In such cases, an alternative to the
traditional approach for choosing the dominant collective pairs must
be sought. Indeed, such considerations have led to the difficulties
that have persisted for many years in microscopically deriving the
IBM for all but spherical $U(5)$-like nuclei, at
least until the recent work in which the collective surface emerging
from density functional treatments is mapped directly into a boson
space, with no direct consideration of the internal structure of the
collective pairs \cite{Otsuka}.

In the next section  we discuss how  to build more appropriately the
dominant collective pairs for our study of the shape transition from
vibrational to deformed nuclei using the NPA, and then in the
results that follow in Section IV we show that the choice of this
new strategy for building the collective pairs, a strategy borrowed
from an earlier effort to build the relevant collective pairs in
deformed nuclei, is critical to obtain a meaningful description of
the shape transition from vibrational to deformed nuclei in the
region we study, namely the even-mass Ce isotopes.

\section{Microscopic structure of collective nucleon pairs}\label{hfb}

As we will see in next section, it is critical to use collective
pairs that are tuned to the dynamics of the collective region under
investigation. In the vibrational region, the traditional pairs that
have been used in the NPA are most likely appropriate as they follow
from the dynamics of generalized seniority. In deformed nuclei,
however, such pairs do not incorporate any effects of the
proton-neutron interaction. Thus, we propose to use a prescription
first discussed by Pittel and Dukelsky \cite{PD} in the context of
the microscopic IBM for deformed nuclei. Their prescription derives
from the use of Hartree-Fock-Bogolyubov (HFB) approximation to
variationally generate an optimum intrinsic pair, from which pairs
of definite angular momentum can then be projected. Similar
considerations to obtain coherent  $SD$ nucleon pairs appropriate to
the NPA calculations based on the field approximation were studied
in Ref. \cite{Maglione}.

We begin the procedure by carrying out an axially-symmetric HFB
calculation in the same space and with the same hamiltonian that is
used in the  NPA study. Once the HFB calculations have been carried
out, we then transform to the canonical basis, namely the basis in
which the density matrix $\rho^{(m)}(\tau)$ for particles of type
$\tau ~(\nu$ or $\pi)$ and projection $m$, is diagonal. Letting
$\tau^{\dagger}_{jm}$ denote the operator that creates a nucleon of
type $\tau$  in orbit $(jm)$, we can express the single-particle
creation operators in the canonical basis by
\begin{equation}
\tau^{\dagger}_{\alpha m} = \sum_{j} C^{(m)}_{\alpha
j}(\tau)~\tau^{\dagger}_{jm}~,
\end{equation}
where $C^{(m)}_{\alpha j}(\tau)$ is the matrix that transforms the
density matrix to diagonal form for projection $m$ and particle type
$\tau$.  The intrinsic HFB wave function in the canonical basis can
be expressed as
\begin{equation}
|\Phi> = c ~ {\rm exp} \left( \sum_{\tau} \Gamma^{\dagger}_{\tau}
\right) ~|~\tilde{0} \rangle ~,
\end{equation}
where
\begin{equation}
\Gamma^{\dagger}_{\tau} =d_{\tau} \sum_{\alpha m>0}
\frac{v^{(\tau)}_{\alpha m}}{u^{(\tau)}_{\alpha m}} ~
\tau^{\dagger}_{\alpha m} \tau^{\dagger}_{\alpha -m}  \label{Gamma}
\end{equation}
is a coherent pair creation operator and $|\tilde{0} \rangle$
denotes the doubly-magic core.

The quantities entering (\ref{Gamma}) can be obtained directly from
the diagonal elements of the density matrix, viz:
\begin{eqnarray*}
v^{(\tau)}_{\alpha m} &=& \left[ \rho^{(m)}_{\alpha \alpha}
\right]^{1/2} ~,\nonumber \\
u^{(\tau)}_{\alpha m} &=& \left[ 1- (v^{(\tau)}_{\alpha m})^2
\right]^{1/2} ~, \nonumber \\
d_{\tau} &=& \left( \sum_{\alpha m>0} \left[v^{(\tau)}_{\alpha
m}/u^{(\tau)}_{\alpha m} \right]^2 \right)^{-1/2}  ~.   \nonumber
\end{eqnarray*}

The correlated pair creation operators $\Gamma^{\dagger}_{\tau}$
contain components with all allowed even angular momenta,
\begin{equation}
\Gamma^{\dagger}_{\tau} =\sum_{L} a^{(\tau)}_L
\Gamma^{{(L)}^{\dagger}}_{\tau}~.  \nonumber
\end{equation}
The square of the amplitude $a^{(\tau)}_L$ determines the importance
of the correlated pair of type $\tau$ with angular momentum $L$ in
the intrinsic wave function and is given by
\begin{eqnarray*}
a^{(\tau)}_L &=& d_{\tau} \left( \sum_{j_1 \leq j_2} \left[
A^{(L)}_{j_1j_2} (\tau) \right] \right)^{1/2} ~, \nonumber \\
A^{(L)}_{j_1j_2} (\tau) &=& \sum_{\alpha m>0}
\left(v^{(\tau)}_{\alpha m}/u^{(\tau)}_{\alpha m}
\right)~C^{(m)}_{\alpha j_1}(\tau) C^{(m)}_{\alpha j_2}(\tau)
  \nonumber \\
&~&\times (-)^{j_2-m} \frac{ \left(1+(-)^L
\right) }{ \sqrt{2}}  \left( j_1 j_2 m-m|L0 \right)~.  \\
\end{eqnarray*}
The structure of the correlated pair
$\Gamma^{{(L)}^{\dagger}}_{\tau}$ with angular momentum $L$ is given
by
\begin{eqnarray}
&& \Gamma^{{(L)}^{\dagger}}_{\tau} = \frac{d_{\tau}}{a_{L}^{(\tau)}}
\sum_{j_1 \leq j_2}~A^{(L)}_{j_1j_2} (\tau) ~\left[
\tau^{\dagger}_{j_1} \tau^{\dagger}_{j_2} \right]^{(L)}
/\sqrt{1+\delta_{j_1j_2}} ~,  \nonumber
\end{eqnarray}
and is the information of interest for use in the NPA when applied to well-deformed nuclei.

\section{Parameters in our NPA calculations}\label{ham}

We treat the even-mass Ce isotopes by assuming a $Z=50$, $N=82$
doubly-magic core and distributing the remaining protons over the
orbits of the $50-82$ major shell and the remaining neutrons over
the orbits of the $82-126$ major shell.

The hamiltonian, as dictated by the computational structure of the
NPA code, takes the form
\begin{eqnarray}
&& H =\sum_{\tau=\nu,\pi } \left( \sum_j \epsilon_{\tau j}
\tau^{\dagger}_{j}\tau_{j} - \sum_{L=0,2}G_{\tau L} A_{\tau}^{\dagger(
L)} \cdot \tilde{A}_{\tau}^{(L)} \right)  \nonumber \\
&& ~~~ - \kappa Q_{\pi} \cdot Q_{\nu} ~.
\end{eqnarray}
It includes a single-particle energy term and a two-body interaction
consisting of multipole pairing interactions in the $nn$ and $pp$
channels and a separable quadrupole-quadrupole interaction in the
$pn$ channel.

\begin{table}
\caption{Single-particle energies (in MeV) used in the calculations for the Ce
isotopes described in the text. They are  based on low-lying energy levels of
$^{133}$Sb and $^{133}$Sn. }\label{sp}
\begin{tabular}{ c c c c c c c }
\hline\hline
  $l_{j_{\pi}}$  & $s_{1/2}$ & $d_{3/2}$ & $d_{5/2}$ & $g_{7/2}$  &   $h_{11/2}$  & \\
 $\epsilon_{\pi}$ & 2.990 & 2.708 & 0.962 & 0.000 & 2.793 & \\
& & & & &  & \\
 $l_{j_{\nu}}$ &   $p_{1/2}$ & $p_{3/2}$ & $f_{5/2}$ & $f_{7/2}$ & $h_{9/2}$ & $i_{13/2}$   \\
 $\epsilon_{\nu}$  & 1.656 & 0.854 & 2.005 & 0.000 & 1.561 &1.800  \\
\hline\hline
\end{tabular}
\end{table}

The single-particle energies are extracted from the spectra of  $^{133}$Sb and $^{133}$Sn,
and are listed in Table \ref{sp}. The two-proton interaction parameters are obtained from the binding
energy and $2_1^+$ energy of $^{134}$Te. The values that emerge are
\begin{eqnarray}
G_{\pi 0} = -0.18~{\rm MeV}, ~~ G_{\pi 2} = 0 ~. \nonumber
\end{eqnarray}
Note that a pure monopole pairing interaction reproduces both the
binding energy of $^{134}$Te {\it and} the excitation energy of the
$2_1^+$ state. The two-neutron parameters are obtained from the binding energy and
$2_1^+$ energy of $^{134}Sn$. The resulting values are
\begin{eqnarray}
G_{\nu 0} = -0.13~{\rm MeV}, ~~ G_{\nu 2} = -0.012~{\rm MeV} ~.   \nonumber
\end{eqnarray}
The strength of the proton-neutron quadrupole interaction is
obtained by requiring an optimum description of both the binding
energy and the lowest $2^+$ excitation energy of $^{136}$Te, which
has two valence protons and two valence neutrons, and is given by
\begin{eqnarray}
\kappa = -0.20~ {\rm MeV}~.  \nonumber
\end{eqnarray}

\section{Calculated results}\label{res}

In this section, we present our calculated results obtained using
the NPA for the even-mass Ce isotopes from $^{142}$Ce  through
$^{148}$Ce. That this series of isotopes exhibits a shape transition
from vibrational to rotational nuclei can be seen from Table
\ref{r}, where we present the ratios of excitation energies found
experimentally for the $4_1^+$ and $6_1^+$ levels relative to the
$2_1^+$ levels through this chain. The nucleus $^{142}$Ce, with
$N_{\nu}=2$ valence neutrons, has a vibrational character, with a
$R_{4/2}$ ratio of $1.90$, while $^{148}$Ce   with $N_{\nu}=8$
valence neutrons has a fairly rotational character with $R_{4/2}=
2.87$ and $R_{6/2}=5.32.$  The transition from vibrational to
rotational is fairly smooth.

\begin{table}
\caption{Ratios of experimental excitation energies of the  Ce  isotopes from Ref.
\cite{exp}. $R_{4/2}$ gives the ratio of the $4_1^+$ to the $2_1^+$ excitation
energies, and $R_{6/2}$ the ratio of the $6_1^+$ to $2_1^+$
energies. The last column, denoted $SU(3)$, gives the results in the rotational
limit.}\label{r}
\begin{tabular}{l c cc c c c c c}
\\
\hline\hline
&&& $^{142}$Ce  & $^{144}$Ce & $^{146}$Ce & $^{148}$Ce & SU(3)&\\
\hline
&$R_{4/2}$ &$~~~$& 1.90& 2.36& 2.58& 2.87 & 3.33&\\
&$R_{6/2}$ && 2.72 & 4.15& 4.52& 5.32 & 7.00 &\\

\hline\hline
\end{tabular}
\end{table}

\begin{figure}
\includegraphics[width = 0.4\textwidth]{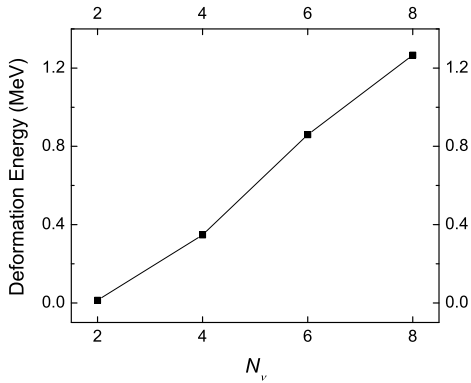}
\caption{ Deformation energies ($E_{\rm deformed}-E_{\rm spherical})$ from
axially-symmetric HFB calculations for the even-mass Ce isotopes as a function
of the number of valence neutrons $N_{\nu}$}\label{de}
\end{figure}

In Fig. \ref{de}, we show the deformation energies $(E_{\rm
deformed}-E_{\rm spherical})$ that derive from our HFB calculations
of these nuclei. As expected, the deformation energy grows gradually
with increasing neutron number, with the gain in energy reaching a
value of $1.27$~MeV in $^{148}$Ce. This is not as large a
deformation energy as we would have liked to have seen for a
well-deformed nucleus, perhaps reflecting the fact that our model
space does not include all the orbits that could contribute
meaningfully. Nevertheless, the results are suggestive that the HFB
calculations are producing the observed shape transition.

\begin{figure}
\includegraphics[height=.35\textwidth]{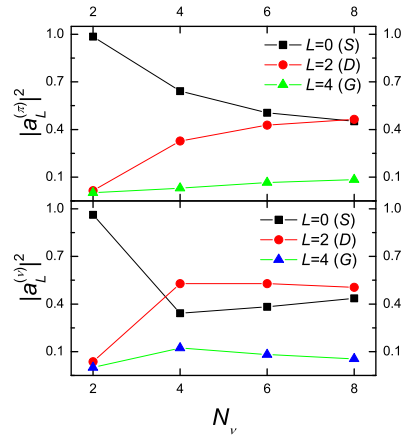}
\caption{Contributions of pairs with definite angular momentum to the HFB
intrinsic states calculated for the Ce isotopes with different neutron pair number $N_{\nu}$.}\label{al}
\end{figure}

In Fig. \ref{al}, we present the calculated HFB results for $|a_L|^2$ for these
isotopes, illustrating the gradually increasing importance of $L \neq 0$ pairs
in the ground band as the deformation grows. By the time we reach $^{148}$Ce,
$D$ pairs contribute to the intrinsic state of the ground band as much as $S$
pairs.

\begin{figure*}
\includegraphics[height=0.9\textwidth]{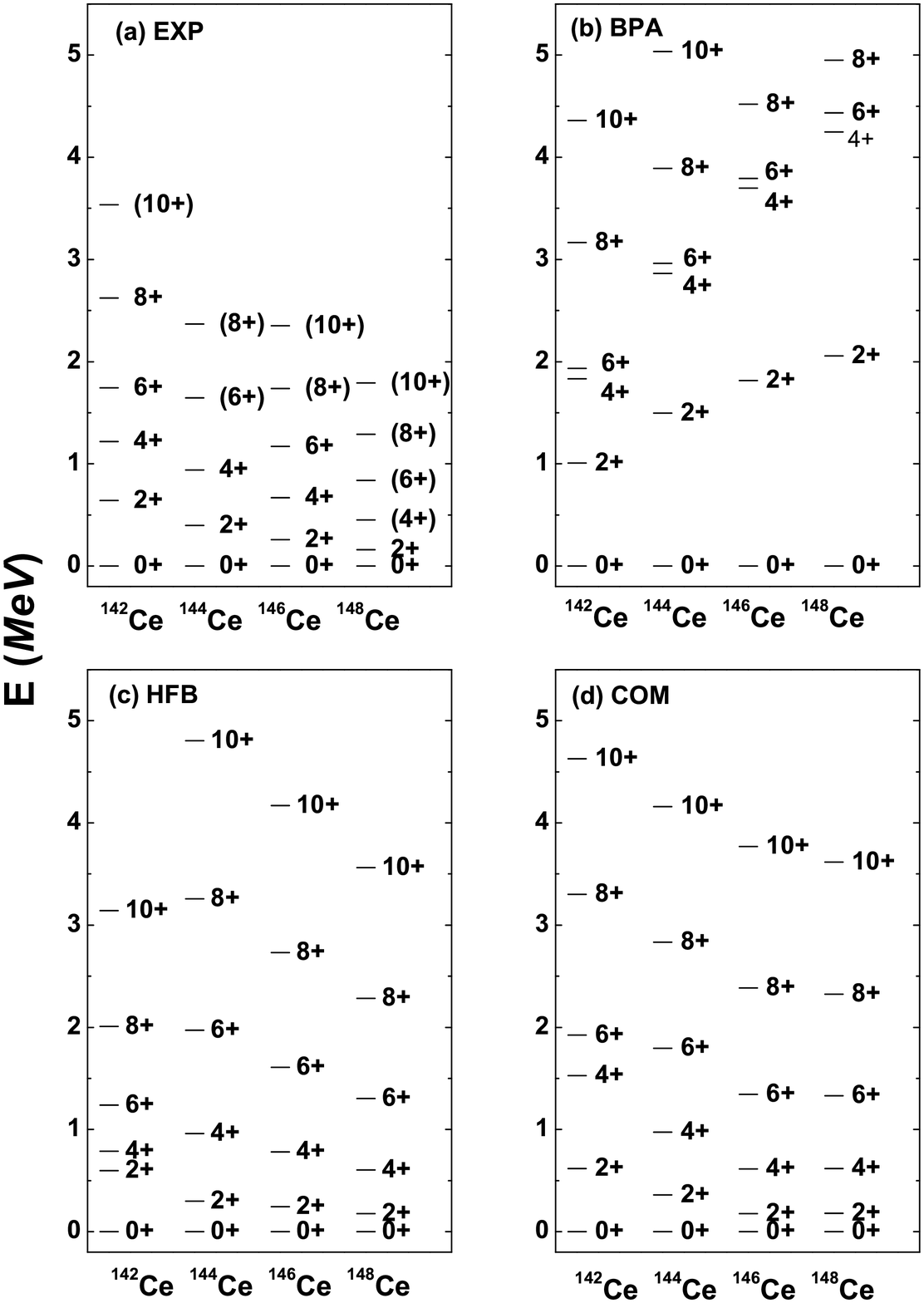}
\caption{ Comparison of the spectra calculated using the NPA  for
the even Ce isotopes (panels b-d) with the corresponding
experimental spectra (panel a, denoted by ``EXP"). The model spaces
are constructed using generalized seniority-two states for both
protons and neutrons (denoted by ``BPA"), and  $SDG$-nucleon pairs
obtained from HFB calculations (denoted by ``HFB"), in panels (b)
and (c), respectively. The model space of panel (d) is a direct sum
of the two bases in panels (b) and (c), and is denoted ``COM". See
the text for details. }\label{spe}
\end{figure*}

Next we present the results of our calculations using the NPA. In
all calculations our model space includes the lowest $S$ pair, the
lowest $D$ pair and the lowest  $G$ pair (with spin four), both for
neutrons and protons. However, because of computational
considerations we only permit a single $G$ pair for each type of
particle. Fig. \ref{spe}  shows several sets of calculated results
in comparison with the experimental data. The spectra denoted ``BPA"
are based on the use of the traditional broken pair approximation
(BPA) prescription for the $S$, $D$ and $G$ pairs, as described in
Section II. The spectra denoted ``HFB" are based on the use of pairs
obtained from axially-symmetric HFB calculations, as described in
Sec. \ref{hfb}. The spectra denoted ``COM" include the $S$ and $D$
pairs from the two prescriptions in the same calculation, with their
nonorthogonality taken into account. We return later to our reason
for including the latter results. Results are shown up to $5.5$ MeV
in excitation energy, both for the experimental and calculated
spectra.

There are several clear messages that emerge from these results.
First, the traditional broken pair approximation (BPA) results are
unable to describe the shape transition from vibrational to
rotational as we proceed through the Ce chain. This is not
surprising, considering our earlier motivating remarks. The HFB
results, in contrast, do give rise to a clear shape transition.
Though $^{142}$Ce does not exhibit a rotational pattern, the
addition of just two neutrons leads to a fairly robust rotational
spectrum, which becomes even more pronounced with the addition of
further neutrons.

\begin{figure}
\includegraphics[height=0.35\textwidth]{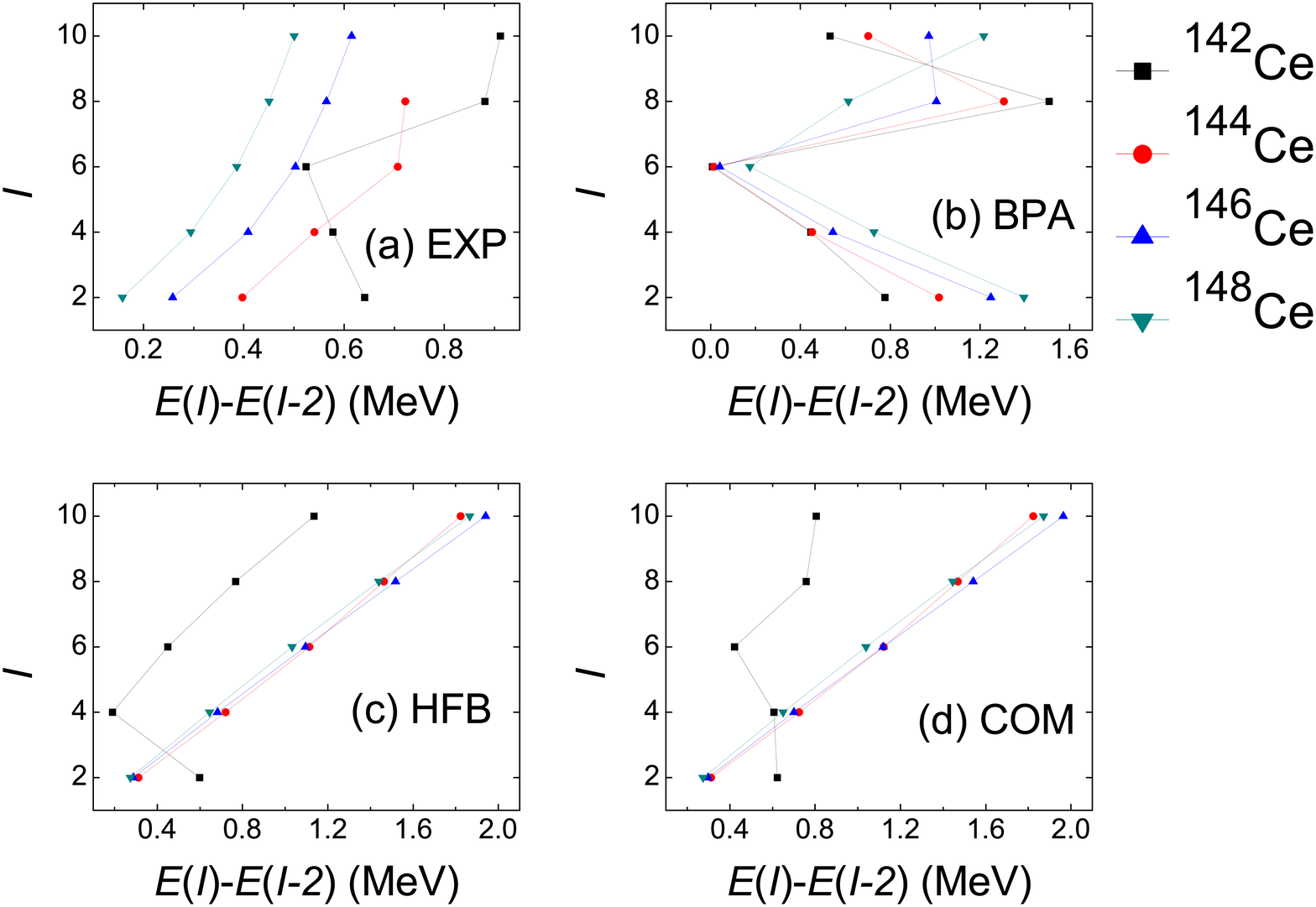}
\caption{Comparison of associated gamma rays [$E(I)-E(I-2)$]
emerging from the three sets of calculations with the corresponding
experimental data. Panels (a-d) have the same meaning as in Fig. 3.
}\label{ray}
\end{figure}

These conclusions are made even more striking when we show the energies of the
associated gamma rays emerging from the two sets of calculations, which are
given in Fig. \ref{ray}. The HFB results show a rapid emergence of a deformed
pattern (roughly parallel lines) with increasing neutron number, whereas none
is seen in the BPA spectra.

Interestingly, the BPA results do not even give a good description
of $^{144}$Ce, with only four valence neutrons. The excitation
energy it predicts for the lowest $2^+$ state is significantly
larger than that seen experimentally. This is to be contrasted with
the HFB results, which already gets a fairly good description of the
lowest $2^+$ excitation energy in $^{144}$Ce.

\begin{figure}
\includegraphics[height=0.35\textwidth]{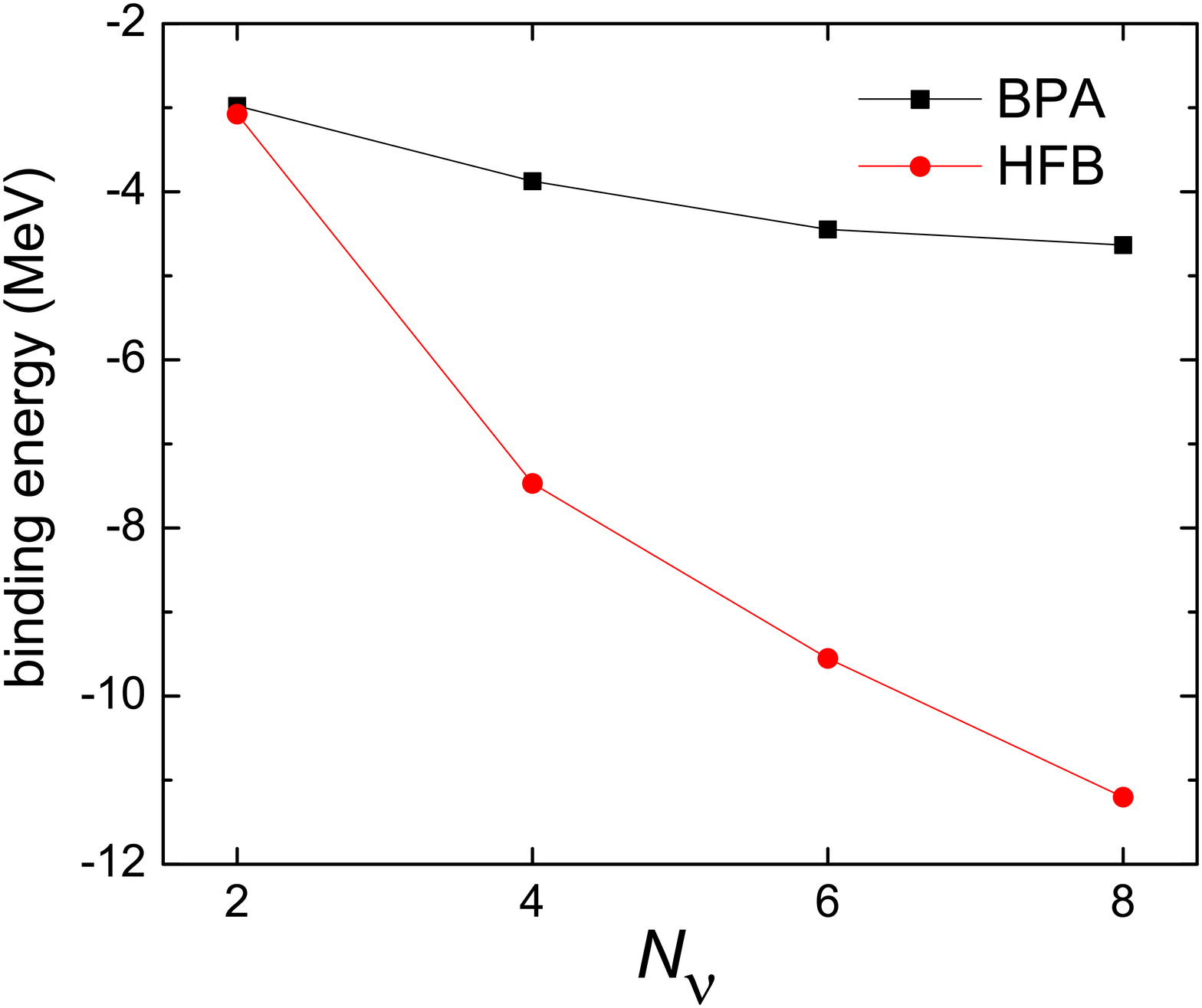}
\caption{ Binding energies arising from the NPA
calculations for the even Ce isotopes as described in the text. The
curve denoted BPA refers to the use of the broken pair approximation
to define the collective pairs; the curve denoted HFB refers to the
use of the Hartree-Fock-Bogolyubov approximation to generate them.} \label{be}
\end{figure}

While we have focussed on energy spectra in Figs. \ref{spe} and
\ref{ray}, and in the discussion that followed, we can get a clearer
picture of why it is important to use dynamically-determined
collective pairs by looking at the absolute binding energies  of the
ground states that emerge in these calculations with respect to the
$^{132}$Sn core. These results are shown in Fig. \ref{be}. When we
use the broken pair approximation prescription for the collective
pairs of the model we obtain much less binding than with the HFB
pairs, which include effects of the proton-neutron interaction in
their structure. While this is especially true for the most
well-deformed nuclei, it is also true, albeit to a much less extent,
in $^{142}$Ce, where the use of HFB collective pairs produces
roughly $0.1$~MeV in additional binding.

While the method is able to produce a shape transition from vibrational to
rotational nuclei, it produces it much too rapidly. As is evident both from
Fig. \ref{spe} and even more so from Fig. \ref{ray}, the model already produces
a fairly well-deformed spectrum for $^{144}$Ce, which according to Table
\ref{r} is clearly transitional.

It is not too surprising that we are unable to adequately describe transitional
nuclei, since our axially-symmetric HFB solution is not very meaningful for
those nuclei. The fairly small HFB deformation energy for those nuclei, as seen
in the Fig. \ref{de}, suggests that the HFB solution is not stable against
zero-point fluctuations. One possible thought is that we might include more
than one pair for each angular momentum in this region. As a first step in this
direction, we have carried out calculations in which we include both the HFB
pairs and the BPA pairs, taking into account their nonorthogonality, to see
whether this improves our description of the transitional region. Those are the
results denoted ``COM" in both Figs. \ref{spe} and \ref{ray}. The inclusion of
a second $S$ and $D$ pair for each type of nucleon improves the description
substantially, giving a better reproduction of the properties of all nuclei
considered. It markedly improves upon the HFB results for $^{142}$Ce, placing
the $2_1^+$ state much closer to its experimental location. However, as is
evident from Fig. \ref{ray}, the shape transition is still too rapid.
Nevertheless, we believe that the idea of including more than one pair with
each angular momentum is a possible approach for future consideration, as it
can be readily included in the analysis, but further thought as to which pairs
to include, especially in the transitional region, is needed.

It is also interesting to note that even in the most deformed
nucleus considered, $^{148}$Ce, the moment of inertia that derives
from the best NPA calculations, those denoted ``COM", is still too
small and the spectrum is accordingly too spread-out. Interestingly
the same result emerged in the first efforts to microscopically
derive the IBM in deformed nuclei by mapping the fermionic density
functional onto a corresponding one in the boson space. Recently it
was shown that it is necessary to add a further term to the IBM
hamiltonian of the form $\alpha J(J+1)$ to better reproduce the
moment of inertia of the rotational systems that were studied in
Ref. \cite{Otsuka2}. The NPA seems to also have need for an
additional contribution to the rotational moment of inertia, which
perhaps has a similar origin. It is not clear at this time, however,
how to incorporate it in the NPA formalism.

\section{Summary and closing remarks}\label{sum}

In this paper, we have addressed the issue of whether or not the NPA
is able to describe the transition from vibrational to rotational
nuclei in real nuclear systems. We focus on the even-mass Ce
isotopes, in which such a shape transition is known to occur
experimentally. We see that with the use of the traditional
collective pairs that have been used in previous applications of the
model, namely pairs derived from the broken pair or generalized
seniority approaches, the model is not able to describe such a shape
transition. Such pairs leave out important effects of the
proton-neutron interaction and thus are not suitable in a truncated
description of collective nuclei with active neutrons {\it and}
protons. In contrast, a description that uses collective pairs from
the HFB approximation does contain the requisite correlation effects
and is able to produce the observed shape transition. On the other
hand, the shape transition produced is too sharp, reflecting the
fact that the correlated pairs produced in the HFB calculations are
not adequate for the transitional region. Furthermore, the moments
of inertia are not very well reproduced.

The difficulty in obtaining appropriate collective pairs to use in the NPA in
the various nuclear regions is reminiscent of the same difficulty that arose in
early efforts to microscopically derive the IBM for these different regions. It
is only with the recent development of methods that map the collective surface
{\it directly} onto an associated IBM collective surface that it has been
possible to consistently obtain a microscopic derivation of the IBM. In the
absence of an analogous procedure for the NPA, we have suggested the
possibility of using several collective pairs in those regions in which a
single HFB pair for each angular momentum is not sufficient. Further work to
identify how to optimally obtain those pairs in transitional regions is still
needed. Likewise, we have seen that the moment of inertia that results from our
best calculation for well deformed nuclei is too small and further work is
needed to find a way to increase the rotational moment of inertia in the NPA
framework.

{\bf Acknowledgements:} The work reported herein was carried out while two of
the authors Y. L. and G. J. F. were visiting the Bartol Research Institute and
the Department of Physics and Astronomy at the University of Delaware, whose
hospitality during this visit they gratefully acknowledge. The work of S. P.,
Y. L. and G. J. F. was supported in part by the National Science Foundation
under Grant No. PHY-0854873,  and that of Y. M. Z., Y. L., and G. J. F. by the
National Science Foundation of China under Grants No. 10975096 and 11145005,
and by the Science \& Technology Committee of   Shanghai city under Grant No.
11DZ2260700.


\section*{References}


\begin{thebibliography}{8}


\bibitem{IBM} F. Iachello and A. Arima, {\it the Interacting Boson Model}  (Cambridge University Press,
  Cambridge, 1987).

\bibitem{phase} P.  Cejnar, J.  Jolie, and R.  F. Casten, Rev. Mod. Phys. {\bf 82}, 2155 (2010).

\bibitem{Otsuka}K. Nomura, N. Shimizu and T. Otsuka, Phys. Rev. Lett. {\bf 101}, 142501 (2008).

\bibitem{Chen} J. Q. Chen, Nucl. Phys. A {\bf 626}, 686 (1997).

\bibitem{NPSM} Y. M. Zhao, N. Yoshinaga, S. Yamaji, J. Q. Chen, and
A. Arima, Phys. Rev. C {\bf 62}, 014304 (2000).

\bibitem{Draayer} Y. Luo, Y. Zhang, X. Meng, F. Pan and J. P. Draayer, Phys. Rev. C {\bf 80}, 014311 (2009).

\bibitem{GeneralizedSeniority} I. Talmi, Nucl. Phys. {\bf A 172}, 1 (1971); Simple Models of Complex Nuclei
 (Harwood, New York, (1993).

\bibitem{BPA} K. Allart, E. Boeker, G. Bonsignori, M. Savoia, and Y. K. Gambhir, Phys. Rep. 169, 209 (1988).

\bibitem{PD} S. Pittel and J. Dukelsky, Phys. Lett. {\bf  B 128}, 9 (1983).

\bibitem{Maglione} E. Maglione, F. Catara, A. Insolia, and A. Vitturi, Nucl. Phys. {\bf A 397}, 102 (1982);
E. Maglione and A. Vitturi, R. A. Broglia, and C. H. Dasso, Nucl. Phys. {\bf A 404}, 333 (1983);
E. Maglione and A. Vitturi, Prog. Part. Nucl. Phys. {\bf 9}, 87 (1983).


\bibitem{exp} Evaluated Nuclear Structure Data, http://www.nndc.bnl.gov/ensdf/.

\bibitem{Otsuka2} K. Nomura, T. Otsuka, N. Shimizu and L. Guo, Phys. Rev. C {\bf 83}, 041302 (2011).

\end{thebibliography}
\end{document}